\documentstyle[preprint,aps,floats,tighten]{revtex}

\input{psfig.tex}

\begin{document}

\preprint{\vbox{\hbox{CU-TP-786} 
                \hbox{CAL-616}
                \hbox{hep-ph/9609531}
}}

\title{PARTICLE DARK MATTER\footnote{To appear in the
proceedings of the VIIIth Rencontres de Blois, {\it Neutrinos,
Dark Matter, and the Universe}, June 8--12, 1996, Blois, France.}}

\author{Marc Kamionkowski\footnote{kamion@phys.columbia.edu}}
\address{Department of Physics, Columbia University, 538 West
120th Street, New York, New York~~10027}
\maketitle

\begin{abstract}
Several ideas for new physics beyond the standard model may
provide particle candidates for the dark matter in the Galactic
halo.  The two leading candidates are an axion and
a weakly-interacting massive particle (WIMP), such as the neutralino in
supersymmetric extensions of the standard model.  Several
possibilities for detection of such particles are discussed.
An assessment of the relative merits of various WIMP-detection
techniques is given.  I then review the prospects for improving
our knowledge of the amount, distribution, and nature of the
dark matter in the Universe from future maps of the cosmic
microwave background.
\end{abstract}

\def\VEV#1{\left\langle #1\right\rangle}
\def\fun#1#2{\lower3.6pt\vbox{\baselineskip0pt\lineskip.9pt
  \ialign{$\mathsurround=0pt#1\hfil##\hfil$\crcr#2\crcr\sim\crcr}}}
\def\la{\mathrel{\mathpalette\fun <}}
\def\ga{\mathrel{\mathpalette\fun >}}
\def\gtrsim{\ga}
\def\lesssim{\la}
\def\slashchar#1{\setbox0=\hbox{$#1$}           
   \dimen0=\wd0                                 
   \setbox1=\hbox{/} \dimen1=\wd1               
   \ifdim\dimen0>\dimen1                        
      \rlap{\hbox to \dimen0{\hfil/\hfil}}      
      #1                                        
   \else					
      \rlap{\hbox to \dimen1{\hfil$#1$\hfil}}   
      /                                         
   \fi}                                         %
\def\hatn{{\bf \hat n}}

\section{Introduction}
Almost all astronomers will agree that most of the mass in the
Universe is nonluminous.  The nature of this dark matter remains
one of the great mysteries of science today.  
Dynamics of cluster of galaxies suggest
a universal nonrelativistic-matter density of
$\Omega_0\simeq0.1-0.3$.  If the luminous matter were all
there was, the duration of the epoch
of structure formation would be very short, thereby requiring
(in almost all theories of structure formation) fluctuations in
the microwave background which would be larger than those observed.  
These considerations imply $\Omega_0\gtrsim0.3$ \cite{marccmb}.
Second, if the current value of $\Omega_0$ is of order unity today,
then at the Planck time it must have
been $1\pm10^{-60}$ leading us to believe that $\Omega_0$ is
precisely unity
for aesthetic reasons.  A related argument comes from
inflationary cosmology, which provides the most satisfying
explanation for the smoothness of the microwave background
\cite{inflation}.  To account for this isotropy, inflation must set
$\Omega$ (the {\it total} density, including a cosmological
constant) to unity.

\begin{figure}[htbp]
\centerline{\psfig{file=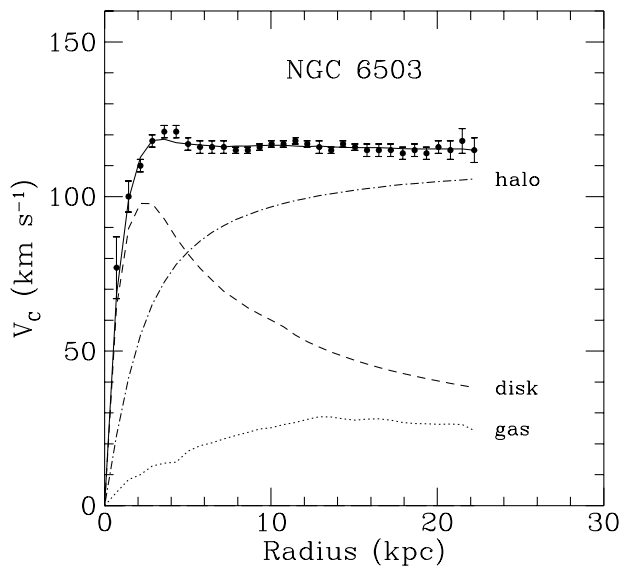,width=3.3in}}
\bigskip
\caption{
         Rotation curve for the spiral galaxy NGC6503.  The points
	 are the measured circular rotation velocities as a
	 function of distance from the center of the galaxy.
	 The dashed and dotted curves are the contribution to
	 the rotational velocity due to the observed disk and
	 gas, respectively, and the dot-dash curve is the
	 contribution from the dark halo.}
\label{rotationfigure}
\end{figure}

However, the most robust observational evidence for the
existence of dark matter involves galactic dynamics.  There is
simply not enough luminous matter ($\Omega_{\rm lum}\la0.01$)
observed in spiral galaxies to account for their observed
rotation curves (for example, that for NGC6503 shown in
Fig.~\ref{rotationfigure} \cite{broeils}).  Newton's
laws imply a galactic dark halo
of mass $3-10$ times that of the luminous component.  With the
simple and plausible assumption that the halo of our galaxy is
roughly spherical, one can determine that the local dark-matter
density is roughly $\rho_0 \simeq 0.3$ GeV~cm$^{-3}$.
Furthermore, the velocity distribution
of the halo dark matter should be roughly Maxwell-Boltzmann with
a velocity dispersion $\simeq270$ km~s$^{-1}$.

On the other hand, big-bang nucleosynthesis
suggests that the baryon density is $\Omega_b\la0.1$ \cite{bbn}, too
small to account for the dark matter in the Universe.
Although a neutrino species of mass ${\cal O}(30\, {\rm eV})$ could
provide the right dark-matter density, N-body simulations of
structure formation in a neutrino-dominated Universe do a poor
job of reproducing the observed structure \cite{Nbody}.
Furthermore, it is difficult to see (essentially from the Pauli
principle) how such a neutrino could make up the dark matter in
the halos of galaxies \cite{gunn}. It appears likely then, that some
nonbaryonic, nonrelativistic matter is required.

The two leading candidates from particle theory are the axion
\cite{axion}, 
which arises in the Peccei-Quinn solution to the strong-$CP$
problem, and a weakly-interacting massive particle (WIMP), which
may arise in supersymmetric (or other) extensions of the
standard model \cite{jkg}.

Here, I review the axion solution to the strong-$CP$ problem, the
astrophysical constraints to the axion mass, and prospects for
detection of an axion.  I then review the WIMP solution to the
dark-matter problem and avenues toward detection.  Finally, I
briefly discuss how measurements of CMB anisotropies may in the
future help determine more precisely the amount of exotic dark
in the Universe.

\section{Axions}

Although supersymmetric particles seem to get more attention in
the literature lately, we should not forget that the axion also
provides a well-motivated and promising alternative dark-matter
candidate \cite{axion}.  The QCD Lagrangian may be written
\begin{equation}
     {\cal L}_{QCD} = {\cal L}_{\rm pert} + \theta {g^2 \over 32
     \pi^2} G \widetilde{G},
\end{equation}
where the first term is the perturbative Lagrangian responsible
for the numerous phenomenological successes of QCD.  However,
the second term (where $G$ is the gluon field-strength tensor
and $\widetilde{G}$ is its dual), which is a consequence of
nonperturbative effects, violates $CP$.  However, we know
experimentally that $CP$ is not violated in the strong
interactions, or if it is, the level of strong-$CP$ violation is
tiny.  From constraints to the neutron electric-dipole moment,
$d_n \lesssim 10^{-25}$ e~cm, it can be inferred that $\theta
\lesssim 10^{-10}$.  But why is $\theta$ so small?  This is the
strong-$CP$ problem.

The axion arises in the Peccei-Quinn solution to the strong-$CP$
problem \cite{PQ}, which close to twenty years after it was proposed still
seems to be the most promising solution.  The idea is to
introduce a global $U(1)_{PQ}$ symmetry broken at a scale
$f_{PQ}$, and $\theta$ becomes a dynamical field which is the
Nambu-Goldstone mode of this symmetry.  
At temperatures below the QCD phase transition,
nonperturbative quantum effects break explicitly the symmetry
and drive $\theta\rightarrow 0$.  The axion is the
pseudo-Nambu-Goldstone boson of this near-global symmetry.  Its
mass is $m_a \simeq\, {\rm eV}\,(10^7\, {\rm GeV}/ f_a)$, and its
coupling to ordinary matter is $\propto f_a^{-1}$.

{\it A priori}, the Peccei-Quinn solution works equally well for
any value of $f_a$ (although one would generically expect it to
be less than or of order the Planck scale).  However, a variety
of astrophysical observations and a few laboratory experiments
constrain the axion mass to be $m_a\sim10^{-4}$ eV, to within a
few orders of magnitude.  Smaller masses would lead to an
unacceptably large cosmological abundance.  Larger masses
are ruled out by a combination of constraints from supernova
1987A, globular clusters, laboratory experiments, and a search
for two-photon decays of relic axions \cite{ted}.

One conceivable theoretical difficulty with this axion mass
comes from generic quantum-gravity arguments \cite{gravity}.  For
$m_a\sim10^{-4}$ eV, the magnitude of the explicit symmetry
breaking is incredibly tiny compared with the PQ scale, so the
global symmetry, although broken, must be very close to exact.
There are physical arguments involving, for example, the
nonconservation of global charge in evaporation of a black hole
produced by collapse of an initial state with nonzero global
charge, which suggest that  global symmetries should be violated
to some extent in quantum gravity.  When one writes down a
reasonable {\it ansatz} for a term in a low-energy effective
Lagrangian which might arise from global-symmetry violation at
the Planck scale, the coupling of such a term is found to be
extraordinarily small (e.g., $\lesssim 10^{-55}$).  Of course,
we have at this point no predictive theory of quantum gravity,
and several mechanisms for forbidding these global-symmetry
violating terms have been proposed \cite{solutions}.  Therefore,
these arguments by no means ``rule out'' the axion solution.
In fact, discovery of an axion would provide much needed clues
to the nature of Planck-scale physics.

Curiously enough, if the axion mass is in the relatively small viable
range, the relic density is $\Omega_a\sim1$ and may therefore
account for the halo dark matter.  Such axions would be produced
with zero momentum by a misalignment mechanism in the early
Universe and therefore act as cold dark matter.  During the process of
galaxy formation, these axions would fall into the Galactic
potential well and would therefore be present in our halo with a
velocity dispersion near 270 km~s$^{-1}$.

Although the interaction of axions with ordinary matter is
extraordinarily weak, Sikivie proposed a very clever method of
detection of Galactic axions \cite{sikivie}.  Just as the axion couples to
gluons through the anomaly (i.e., the $G\widetilde{G}$ term),
there is a very weak coupling of an axion to photons through the
anomaly.  The axion can therefore decay to two
photons, but the lifetime is $\tau_{a\rightarrow \gamma\gamma}
\sim 10^{50}\, {\rm s}\, (m_a / 10^{-5}\, {\rm eV})^{-5}$ which
is huge compared to the lifetime of the Universe and therefore
unobservable.  However, the $a\gamma\gamma$ term in the
Lagrangian is ${\cal L}_{a\gamma\gamma} \propto a {\vec E} \cdot
{\vec B}$ where ${\vec E}$ and ${\vec B}$ are the electric and
magnetic field strengths.  Therefore, if one immerses a resonant
cavity in a strong magnetic field, Galactic axions which pass
through the detector may be converted to fundamental excitations
of the cavity, and these may be observable \cite{sikivie}.  Such
an experiment
is currently underway and expects to probe the entire acceptable
parameter space within the next five years
\cite{axionexperiment}.  A related experiment, which looks for
excitations of Rydberg atoms, may also find dark-matter axions
\cite{rydberg}.
Although the sensitivity of this technique is supposed to be
excellent, it can only cover a limited axion-mass range.

It should be kept in mind that there are no accelerator tests
for axions in the acceptable mass range.  Therefore, these
dark-matter axion experiment are actually our {\it only}
way to test the Peccei-Quinn solution.

\section{Weakly-Interacting Massive Particles}

Suppose that in addition to the known particles of the
standard model, there exists a new, yet undiscovered, stable (or
long-lived) weakly-interacting massive
particle (WIMP), $\chi$.  At temperatures
greater than the mass of the particle, $T\gg m_\chi$, the
equilibrium number density of such particles is $n_\chi \propto
T^3$, but for lower temperatures, $T\ll m_\chi$, the equilibrium
abundance is exponentially suppressed, $n_\chi \propto
e^{-m_\chi/T}$.  If the expansion of the Universe were so slow
that  thermal equilibrium was always maintained, the number of
WIMPs today would be infinitesimal.  However, the Universe is
not static, so equilibrium thermodynamics is not the entire story.

%
\begin{figure}[htbp]
\centerline{\psfig{file=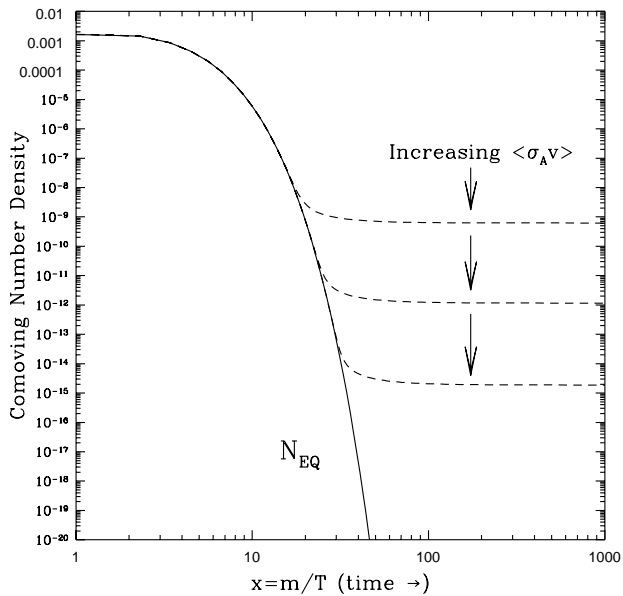,width=3.3in}}
\bigskip
\caption{
        Comoving number density of a WIMP in the early
	Universe.  The dashed curves are the actual abundance,
	and the solid curve is the equilibrium abundance.}
\label{YYY}
\end{figure}

At high temperatures ($T\gg m_\chi$), $\chi$'s are abundant and
rapidly converting to lighter particles and {\it vice versa}
($\chi\bar\chi\leftrightarrow l\bar l$, where $l\bar l$ are quark-antiquark and
lepton-antilepton pairs, and if $m_\chi$ is greater than the mass of the
gauge and/or Higgs bosons, $l\bar l$ could be gauge- and/or Higgs-boson
pairs as well).  Shortly after $T$ drops below $m_\chi$ the number
density of $\chi$'s drops exponentially, and the rate for annihilation of
$\chi$'s, $\Gamma=\VEV{\sigma v} n_\chi$---where $\VEV{\sigma v}$ is the
thermally averaged total cross section for annihilation of $\chi\bar\chi$
into lighter particles times relative velocity $v$---drops below 
the expansion rate, $\Gamma\la H$.  At this point, the $\chi$'s cease to
annihilate, they fall out of equilibrium, and a relic cosmological
abundance remains.

Fig.~\ref{YYY} shows numerical solutions to the Boltzmann
equation which determines the WIMP abundance.  The
equilibrium (solid line) and actual (dashed lines) abundances
per comoving volume are plotted as a
function of $x\equiv m_\chi/T$ (which increases with increasing time).
As the annihilation cross section is increased
the WIMPs stay in equilibrium longer, and we are left with a
smaller relic abundance.

An approximate solution to the Boltzmann equation yields the
following estimate for the current cosmological abundance of the
WIMP:
\begin{equation}
     \Omega_\chi h^2={m_\chi n_\chi \over \rho_c}\simeq
     \left({3\times 10^{-27}\,{\rm cm}^3 \, {\rm sec}^{-1} \over
     \sigma_A v} \right),
\label{eq:abundance}
\end{equation}
where $h$ is the Hubble constant in units of 100
km~s$^{-1}$~Mpc$^{-1}$.  The result is to a first approximation
independent of the WIMP mass and is fixed primarily by its
annihilation cross section.

The WIMP velocities at freeze out are typically some appreciable
fraction of the speed of light.  Therefore, from
equation~(\ref{eq:abundance}), the WIMP will have a cosmological
abundance of order unity today if the annihilation cross section
is roughly $10^{-9}$ GeV$^{-2}$.  Curiously, this is the order
of magnitude one would expect from a typical electroweak cross
section, 
\begin{equation}
     \sigma_{\rm weak} \simeq {\alpha^2 \over m_{\rm weak}^2},
\end{equation}
where $\alpha \simeq {\cal O}(0.01)$ and $m_{\rm weak} \simeq
{\cal O}(100\, {\rm GeV})$.  The value of the cross section in
equation~(\ref{eq:abundance}) needed to provide $\Omega_\chi\sim1$
comes essentially from the age of the Universe.  However, there
is no {\it a priori} reason why this cross section should be of
the same order of magnitude as the cross section one would
expect for new particles with masses and interactions
characteristic of the electroweak scale.  In other words, why
should the age of the Universe have anything to do with
electroweak physics?  This ``coincidence'' suggests that if a
new, yet undiscovered, massive particle with electroweak
interactions exists, then it should have a relic density of
order unity and therefore provides a natural dark-matter
candidate.  This argument has been the driving force behind a
vast effort to detect WIMPs in the halo.

The first WIMPs considered were massive Dirac or Majorana
neutrinos with masses in the range of a few GeV to a few TeV.
(Due to the Yukawa coupling which gives a neutrino its mass, the
neutrino interactions become strong above a few TeV, and it no
longer remains a suitable WIMP candidate \cite{unitarity}.)  LEP ruled out
neutrino masses below half the $Z^0$ mass.  Furthermore, heavier
Dirac neutrinos have been ruled out as the primary component of
the Galactic halo by direct-detection experiments (described
below) \cite{heidelberg}, and heavier Majorana neutrinos have
been ruled out by indirect-detection experiments
\cite{kamiokande} (also described below) over much
of their mass range.  Therefore, Dirac neutrinos cannot comprise
the halo dark matter \cite{griestsilk}; Majorana neutrinos can,
but only over a
small range of fairly large masses.  This was a major triumph
for experimental particle astrophysicists:\ the first
falsification of a dark-matter candidate.  However, theorists
were not too disappointed:  The stability of a fourth generation
neutrino had to be postulated {\it ad hoc}---it was not
guaranteed by some new symmetry.  So although heavy neutrinos
were plausible, they certainly were not very well-motivated from
the perspective of particle theory.

A much more promising WIMP candidate comes from supersymmetry
(SUSY) \cite{jkg,haberkane}.  SUSY was
hypothesized in particle physics to cure the naturalness problem
with fundamental Higgs bosons at the electroweak scale.
Coupling-constant unification at the GUT scale seems to be
improved with SUSY, and it seems to be an essential ingredient
in theories which unify gravity with the other three fundamental
forces.

As another consequence, the existence of a new symmetry,
$R$-parity, in SUSY theories guarantees that the lightest
supersymmetric particle (LSP) is stable.
In the minimal supersymmetric extension of the
standard model (MSSM), the LSP is usually the neutralino, a linear
combination of the supersymmetric partners of the photon, $Z^0$,
and Higgs bosons.  (Another possibility is the sneutrino, but
these particles interact like neutrinos and have been ruled out
over most of the available mass range \cite{sneutrino}.)  Given
a SUSY model, the cross section for
neutralino annihilation to lighter particles is straightforward,
so one can obtain the cosmological mass density.  The
mass scale of supersymmetry must be of order the weak scale to
cure the naturalness problem, and the neutralino will have only
electroweak interactions.  Therefore, it is to be expected that
the cosmological neutralino abundance is of order unity.  In
fact, with detailed calculations, one finds that the neutralino
abundance in a very broad class of supersymmetric extensions of
the standard model is near unity and can therefore account for
the dark matter in our halo \cite{ellishag}.

If neutralinos reside in the halo, there are several avenues
for detection \cite{jkg}.  One of the most promising techniques currently
being pursued involves searches for the ${\cal O}({\rm keV})$ recoils
produced by elastic scattering of neutralinos from nuclei in
low-background detectors \cite{witten,labdetectors}.  Another strategy is
observation of energetic neutrinos produced by annihilation of
neutralinos in the Sun and Earth in converted proton-decay and
astrophysical-neutrino detectors (such as MACRO, Kamiokande,
IMB, AMANDA, and NESTOR) \cite{SOS}.  There are also searches for
anomalous cosmic rays which would be produced by
annihilation of WIMPs in the halo.  Of course, SUSY particles
should also show up in accelerator searches if their mass falls
within the experimentally accessible range.

Although supersymmetry provides perhaps the most promising
dark-matter candidate (and solves numerous problems in particle
physics), a practical difficulty with supersymmetry is that we
have little detailed predictive power.  In SUSY models, the
standard-model particle spectrum is more than doubled, and we
really have no idea what the masses of all these superpartners
should be.  There are also couplings, mixing angles, etc.
Therefore, what theorists generally do is survey a huge
set of models with masses and couplings within a
plausible range, and present results for relic abundances and
direct- and indirect-detection rates, usually as scatter plots
versus neutralino mass.  

Energetic neutrinos from WIMP annihilation in the Sun or Earth
would be inferred by observation of neutrino-induced upward
muons coming from the direction of the Sun or the core of the
Earth.  Predictions for the fluxes of such muons in SUSY models
seem to fall for the most part between $10^{-6}$ and 1
event~m$^{-2}$~s$^{-1}$ \cite{jkg}, although the numbers may be a bit
higher or lower in some models.  Presently, IMB and Kamiokande
constrain the flux of energetic neutrinos from the Sun to be
less than about 0.02 m$^{-2}$~s$^{-1}$ \cite{kamiokande,imb}.
MACRO expects to be
able to improve on this sensitivity by perhaps an order of
magnitude.  Future detectors may be able to improve on this
limit further.  For example, AMANDA expects to have an area of
roughly $10^4$ m$^2$, and a $10^6$-m$^2$
detector is being discussed.  However, it should be kept in
mind that without muon energy resolution, the sensitivity of
these detectors will not approach the inverse exposure; it will
be limited by the atmospheric-neutrino background.  If a
detector has good angular resolution, the signal-to-noise ratio
can be improved, and even moreso with energy resolution, so
sensitivities approaching the inverse exposure could be
achieved \cite{joakim}.  Furthermore, ideas for neutrino detectors with
energy resolution are being discussed \cite{wonyong}, although
at this point these appear likely to be in the somewhat-distant future.

The other possibility is direct detection of a WIMP via
observation of the nuclear recoil induced by WIMP-nucleus
elastic scattering in a low-background detector.  The predicted
rates depend on the target nucleus adopted.  For example, in a
broad range of SUSY models, the predicted scattering rates in a
germanium detector seem to fall for the most part between
$10^{-4}$ to 10 events~kg$^{-1}$~day$^{-1}$ \cite{jkg}, although again,
there may be models with higher or lower rates.  Current
experimental sensitivities in germanium detectors are around 10
events~kg$^{-1}$~day$^{-1}$ \cite{heidelberg}.  To illustrate
future prospects, consider the CDMS experiment \cite{cdms} which
expects to soon have a kg germanium detector with an 
background rate of 1 event~day$^{-1}$.  After a one-year
exposure, their sensitivity would therefore be ${\cal O}(0.1\,
{\rm event~kg}^{-1}\,{\rm day}^{-1})$; this could be improved
with better background rejection.  Future detectors will
achieve better
sensitivities, and it should be kept in mind that numerous other
target nuclei are being considered by other groups.  However, it
also seems clear that it will be quite a while until a good
fraction of the available SUSY parameter space is probed.

Generally, most theorists have just plugged in SUSY parameters
into the machinery which produces detection rates and plotted
results for direct and indirect detection.  However, another
approach is to compare, in a somewhat model-independent although
approximate fashion, the rates for direct and indirect
detection \cite{jkg,taorich,bernard}.  The underlying
observation is that the rates for the 
two types of detection are both controlled primarily by the WIMP-nucleon
coupling.  One must then note that WIMPs generally undergo one
of two types of interaction with the nucleon: an axial-vector
interaction in which the WIMP couples to the nuclear spin
(which, for nuclei with nonzero angular momentum is roughly 1/2
and {\it not} the total angular momentum), and a scalar
interaction in which the WIMP couples to the total mass of the
nucleus.  The direct-detection rate depends on the WIMP-nucleon
interaction strength and on the WIMP mass.  On the other hand,
indirect-detection rates will have an additional dependence on
the energy spectrum of neutrinos from WIMP annihilation.  By
surveying the various possible neutrino energy spectra, one
finds that for a given neutralino mass and annihilation rate in
the Sun, the largest upward-muon flux is roughly three times as
large as the smallest \cite{bernard}.  So even if we assume the
neutralino-nucleus interaction is purely scalar or purely
axial-vector, there will still be a residual model-dependence of
a factor of three when comparing direct- and indirect-detection
rates.

For example, for scalar-coupled WIMPs, the event rate in a kg
germanium detector will be
equivalent to the event rate in a $(2-6)\times 10^6$ m$^2$
neutrino detector for 10-GeV WIMPs and $(3-5)\times10^4$ m$^2$
for TeV WIMPs \cite{bernard}.  Therefore, the relative
sensitivity of indirect detection when compared with the
direct-detection sensitivity increases with mass.
The bottom line of such an analysis seems to be that
direct-detection experiments will be more sensitive to
neutralinos with scalar interactions with nuclei, although
very-large neutrino telescopes may achieve comparable
sensitivities at larger WIMP masses.  This should
come as no surprise given the fact that direct-detection
experiments rule out Dirac neutrinos \cite{heidelberg}, which
have scalar-like interactions, far more effectively than
do indirect-detection experiments \cite{bernard}.

Generically, the sensitivity of
indirect searches (relative to direct searches) should be better
for WIMPs with axial-vector interactions, since the Sun is
composed primarily of nuclei with spin (i.e., protons).
However, a comparison of direct-
and indirect-detection rates is a bit more difficult for
axially-coupled WIMPs, since the nuclear-physics uncertainties
in the neutralino-nuclear cross section are much greater, and
the spin distribution of each target nucleus must be modeled.
Still, in a careful analysis, Rich and Tao found that in 1994,
the existing sensitivity of energetic-neutrino searches to
axially-coupled WIMPs greatly exceeded the sensitivities of
direct-detection experiments \cite{taorich}.

To see how the situation may change with future
detectors, let us consider a specific axially-coupled
dark-matter candidate, the light Higgsino recently put forward by
Kane and Wells \cite{kanewells}.  In order to explain the
anomalous CDF $ee\gamma
\gamma + \slashchar{E}_T$ \cite{CDF}, the $Z\rightarrow b\bar b$ anomaly,
and the dark matter, this Higgsino must have a mass between
30--40 GeV.  Furthermore, the coupling of this Higgsino to
quarks and leptons is due primarily to $Z^0$ exchange with a
coupling proportional to $\cos 2\beta$, where $\tan\beta$ is the
usual ratio of Higgs vacuum expectation values in
supersymmetric models.  Therefore, the usually messy cross
sections one deals with in a general MSSM simplify for this
candidate, and the cross sections needed for the cosmology of
this Higgsino depend only on the two parameters $m_\chi$ and
$\cos2\beta$.  Furthermore, since the neutralino-quark
interaction is due only to $Z^0$ exchange, this Higgsino will
have only axial-vector interactions with nuclei.

The Earth is composed primarily of spinless nuclei, so WIMPs
with axial-vector interactions will not be captured in the Earth,
and we expect no neutrinos from WIMP annihilation therein.
However, most of the mass in the Sun is composed of nuclei with
spin (i.e., protons).  The flux of upward muons induced by
neutrinos from
annihilation of these light Higgsinos would be $\Gamma_{\rm
det}\simeq 2.7\times10^{-2}\, {\rm m}^{-2}\, {\rm yr}^{-1}\,
\cos^2 2\beta$ \cite{katie}.  On the other hand, the rate for scattering from
$^{73}$Ge is $R\simeq 300\, \cos^2 2\beta\, {\rm kg}^{-1}\,
{\rm yr}^{-1}$ \cite{kanewells,katie}.  For illustration, in
addition to their kg of
natural germanium, the CDMS experiment also plans to
run with 0.5 kg of (almost) purified $^{73}$Ge.  With a
background event rate of roughly one event~kg$^{-1}$~day$^{-1}$,
after one year, the $3\sigma$ sensitivity of the experiment will
be roughly 80 kg$^{-1}$~yr$^{-1}$.  Comparing the predictions
for direct and indirect detection of this axially-coupled WIMP,
we see that the enriched-$^{73}$Ge sensitivity should improve on
the {\it current} 
limit to the upward-muon flux ($0.02$ m$^{-2}$ yr$^{-1}$)
roughly by a factor of 4.  When we compare this with the forecasted
factor-of-ten improvement expected in MACRO, it appears that the
sensitivity of indirect-detection experiments looks more
promising.  Before
drawing any conclusions, however, it should be noted that the
sensitivity in detectors with other nuclei with spin may be
significantly better.  On the other hand, the sensitivity of
neutrino searches increases relative to direct-detection
experiments for larger WIMP masses.  It therefore seems at this
point that the two schemes will be competitive for detection of
light axially-coupled WIMPs, but the neutrino telescopes may
have an advantage in probing larger masses.

A common question is whether
theoretical considerations favor a WIMP which has predominantly
scalar or axial-vector couplings.  Unfortunately, there is no
simple answer.  When detection of supersymmetric dark matter was
initially considered, it seemed that the neutralino in most
models would have predominantly axial-vector interactions.  It
was then noted that in some fraction of models where the
neutralino was a mixture of Higgsino and gaugino, there could be
some significant scalar coupling as well \cite{kim}.  As it became evident
that the top quark had to be quite heavy, it was realized that
nondegenerate squark masses would give rise to scalar couplings
in most models \cite{drees}.  However, there are still large regions of
supersymmetric parameter space where the neutralino has
primarily axial-vector interactions, and in fact, the Kane-Wells
Higgsino candidate has primarily axial-vector interactions.  The
bottom line is that theory cannot currently reliable say which
type of interaction the WIMP is likely to have, so
experiments should continue to try to target both.

\section{Dark Matter and the Cosmic Microwave Background}

The key argument for nonbaryonic dark matter relies on the
evidence that the total nonrelativistic-matter density $\Omega_0
\ga 0.1$, outweighs the baryon density $\Omega_b\la 0.1$ allowed by
big-bang nucleosynthesis.  With the advent of a new
generation of long-duration balloon-borne and ground-based
interferometry experiments and NASA's MAP \cite{MAP} and ESA's
COBRAS/SAMBA \cite{COBRAS} missions, CMB measurements will usher
in a new era in cosmology.  In forthcoming years, the cosmic
microwave background (CMB) may provide a precise inventory of
the matter content in the Universe and confirm the discrepancy
between the baryon density and the total nonrelativistic-matter
density, if it indeed exists.  

The primary goal of these experiments is recovery of the
temperature autocorrelation function or angular power spectrum
of the CMB.  The fractional temperature perturbation
$\Delta T(\hatn)/T$ in a given direction $\hatn$ can be expanded
in terms of spherical harmonics,
\begin{equation}
     {\Delta T(\hatn) \over T} = \sum_{lm} \, a_{(lm)}\,
     Y_{(lm)}(\hatn),
\label{Texpansion}
\end{equation}
where the multipole coefficients are given by
\begin{equation}
     a_{(lm)} = \int\, d\hatn\, Y_{(lm)}^*(\hatn) \, {\Delta
     T(\hatn) \over T}.
\label{alms}
\end{equation}
Cosmological theories predict that these multipole coefficients
are  statistically independent and are distributed with variance
$ \VEV{a_{(lm)}^* a_{(l'm')} } = C_l \, \delta_{ll'} \,
\delta_{mm'}$.
Roughly speaking, each $C_l$ measures the square of the mean temperature
difference between two points separated by an angle $\theta\sim
\pi/l$.

\begin{figure}[htbp]
\centerline{\psfig{file=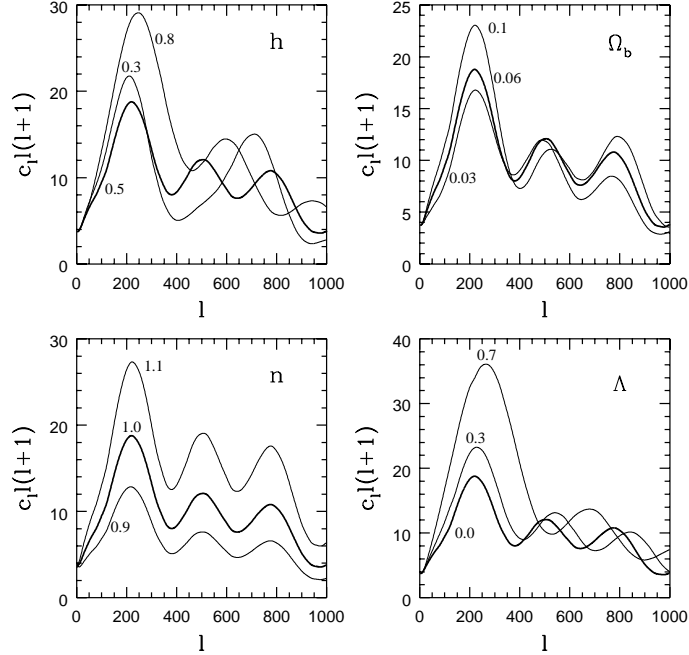,width=3.7in}}
\bigskip
\caption{ 
	  Theoretical predictions for CMB spectra as a function
	  of multipole moment $\ell$ for models with primordial
	  adiabatic perturbations.  In each case, the
	  heavy curve is that for the standard-CDM values,
	  a total density $\Omega=1$, cosmological constant
	  $\Lambda=0$, baryon density $\Omega_b=0.06$, and
	  Hubble parameter $h=0.5$.  Each graph shows the effect
	  of variation of one of these parameters.  In (d),
	  $\Omega=\Omega_0+\Lambda=1$.}
\label{curves}
\end{figure}

Theoretical predictions for the $C_l$'s can be made given a
theory for structure formation and the values of several
cosmological parameters.  For example, Fig.~\ref{curves} shows predictions
for multipole moments in models with primordial adiabatic
perturbations.  
The peaks in the spectra come from oscillations in the
photon-baryon fluid at the surface of last scatter, and the
damping at small angles is due to the finite thickness of the
surface of last scatter.  Each panel shows the
effect of independent variation of one of the cosmological
parameters.  As illustrated, the height, width, and spacing
of the acoustic peaks in the angular spectrum depend on  these
(and other) cosmological parameters.  The CMB
spectrum also depends on the model (e.g., inflation or
topological defects) for structure formation, the ionization
history of the Universe, and the presence of gravity waves.
However, no two of the classical cosmological parameters affects
the CMB spectrum in precisely the same way.  For example, the
angular position of the first peak depends primarily  on the
geometry ($\Omega=\Omega_0 +\Lambda$ where $\Lambda$ is the
contribution of the cosmological constant) of the Universe
\cite{kamspergelsug}, but is relatively insensitive to
variations in the other parameters.  Assuming that the
primordial perturbations were adiabatic, we could fit for all of
these parameters if the angular spectrum could be measured
precisely.

COBE normalizes the amplitude and slope of the CMB spectrum to
$\sim10\%$.  However, the angular resolution was not fine enough to probe
the detailed shape of the acoustic peaks in the
power spectrum, so COBE was unable to capitalize on this wealth
of information.  Nor can it discriminate between scalar and
tensor modes.  A collection of recent ground-based
and balloon-borne experiments seem to confirm
a first acoustic peak, but they still cannot
determine its precise height, width, or location.
In the next few years, long-duration balloon
flights (e.g., BOOMERANG and TOPHAT) and ground-based
interferometry experiments (e.g., CAT and CBI) will begin
to discern the first and higher few peaks.
Subsequently, future satellite experiments, such as NASA's MAP
mission \cite{MAP} and then ESA's COBRAS/SAMBA \cite{COBRAS} will
accurately map the CMB temperature over most of the sky with
good angular resolution and will therefore be able to recover
the CMB power spectrum with precision.  

Of course, the precision attainable is ultimately  limited by
cosmic variance and practically by
a finite angular resolution, instrumental noise, and
partial sky coverage in a realistic CMB mapping experiment.  
Assuming that primordial perturbations are adiabatic, one
finds that with future
satellite missions, $\Omega$ may potentially be determined to
better than 10\% {\it after} marginalizing over all other undetermined
parameters, and better than 1\% if the other
parameters can be fixed by independent observations or
assumption \cite{jkksone}.  This would be far more accurate
than any traditional
determinations of the geometry.  (Of course, if primordial
perturbations turn out to be isocurvature or due to topological
defects, this may not be the case.)  The cosmological
constant $\Lambda$ will be determined with a similar accuracy, so
the nonrelativistic-matter density $\Omega_0$ will also be
accurately determined \cite{jkkstwo}.  Small variations in the
baryon density
have a dramatic effect on the CMB spectrum, so $\Omega_b$ will
be determined with even greater precision.  Therefore, if there
is more nonrelativistic matter in the Universe than the baryons
can account for, as current evidence suggests, it should become
clear with these future CMB experiments.

The CMB will also measure the Hubble constant and perhaps be
sensitive to a small neutrino mass \cite{gds}.  Temperature maps
will also begin to disentangle the scalar and tensor (i.e.,
long-wavelength gravity-wave) contributions to the CMB and
determine their primordial spectra, and this could be used to
test inflation \cite{jkkstwo}.  CMB polarization maps may also help
isolate the tensor contribution \cite{polarization}.  Therefore,
the CMB will become an increasingly powerful probe of the early
Universe.

\acknowledgements{This work was supported by the D.O.E. under
contract DEFG02-92-ER 40699, NASA under NAG5-3091, and the
Alfred P. Sloan Foundation.}

\vfill

\begin{thebibliography}{99}{
\bibitem{marccmb} M. Kamionkowski and D. N. Spergel, {\rm Astrophys.
      J.} {\bf 432}, 7 (1994).

\bibitem{inflation} A. H. Guth, Phys. Rev. {\bf D28}, 347 (1981);
    A. D. Linde, Phys. Lett. {\bf 108B}, 389 (1982); A. Albrecht
    and P. J. Steinhardt, Phys. Rev. Lett. {\bf 48}, 1220
    (1982).

\bibitem{broeils} K. G. Begeman, A. H. Broeils, R. H. Sanders,
     1991, Mon. Not. R. Astr. Soc. {\bf 249}, 523 (1991).

\bibitem{bbn} K. A. Olive, G.~Steigman, D.~N.~Schramm, T.~P.~Walker,
      and H.~Kang, {\rm Astrophys. J.} {\bf 376}, 51 (1991).

\bibitem{Nbody} S. D. M. White, C. S. Frenk, and M. Davis,
     Astrophys. J. {\bf 274}, L1-5 (1983).

\bibitem{gunn} S. Tremaine and J. E. Gunn, {\rm Phys. Rev. Lett.},
     {\bf 42}, 407 (1979).

\bibitem{axion} For reviews, see, e.g., M. S. Turner,
     Phys. Rep. {\bf 197}, 67 (1990); G. G. Raffelt,
     Phys. Rep. {\bf 198}, 1 (1990).  An illuminating
     pedagogical discussion has also been given recently by
     P. Sikivie, hep-ph/9506229, talk given at 30th Rencontres de
     Moriond: Euroconferences: Dark Matter in Cosmology, Clocks
     and Tests of Fundamental Laws, Villars sur Ollon,
     Switzerland, 21-28 Jan 1995.

\bibitem{jkg} G. Jungman, M. Kamionkowski, and K. Griest,
     Phys. Rep. {\bf 267}, 195 (1996).

\bibitem{PQ} R. D. Peccei and H. R. Quinn, Phys. Rev. Lett. {\bf
     38}, 1440 (1977); F. Wilczek, Phys. Rev. Lett. {\bf 40},
     279 (1978); S. Weinberg, Phys. Rev. Lett. {\bf 40}, 223
     (1978).

\bibitem{ted} M. A. Bershady, M. T. Ressell and M. S. Turner,
     Phys. Rev. Lett. {\bf 66}, 1398 (1991).

\bibitem{gravity} M. Kamionkowski and J. March-Russell,
     Phys. Lett. {\bf B282}, 137 (1992); R. Holman et al.,
     Phys. Lett. {\bf B282}, 132 (1992); S. M. Barr and
     D. Seckel, Phys. Rev. {\bf D46}, 539 (1992);
     B. A. Dobrescu, hep-ph/9609221.

\bibitem{solutions} R. Holman et al., Phys. Lett. {\bf B282},
     132 (1992); N. Turok, Phys. Rev. Lett. {\bf 76}, 1015
     (1996); R. Kallosh et al., Phys. Rev. {\bf D52}, 912
     (1995); E. A. Dudas, Phys. Lett. {\bf B325}, 124 (1994);
     K. S. Babu and S. M. Barr, Phys. Lett. {\bf B300}, 367 (1993). 

\bibitem{sikivie} P. Sikivie, Phys. Rev. Lett {\bf 51}, 1415
     (1983). 

\bibitem{axionexperiment} K. van Bibber et al.,
     Int. J. Mod. Phys. Suppl. {\bf D3}, 33 (1994).  

\bibitem{rydberg} S. Matsuki, talk given at the International
     Conference on Sources and Detection of Dark Matter in the
     Universe Santa Monica, CA, February 14--16, 1996.

\bibitem{unitarity} K. Griest and M. Kamionkowski, {\rm
     Phys.~Rev. Lett.} {\bf 64}, 615 (1990).

\bibitem{heidelberg} M. Beck, Nucl. Phys. (Proc. Suppl.) {\bf
     B35}, 150 (1994); M. Beck et al., Phys. Lett. {\bf B336},
     141 (1994); S. P. Ahlen et al., {\rm Phys. Lett.} {\bf
     B195}, 603 (1987); D.~O.~ Caldwell et al., {\rm
     Phys.~Rev.~Lett.} {\bf 61}, 510 (1988).

\bibitem{kamiokande} M.~Mori et al.\ (Kamiokande Collaboration), 
	{\rm Phys.~Lett.} {\bf B289}, 463 (1992); M.~Mori et al.\
	(Kamiokande Collaboration),  {\rm Phys.~Rev.} {\bf D48},
	5505 (1993).

\bibitem{griestsilk} K.~Griest and J.~Silk, \rm Nature \bf
     343\rm, 26 (1990); L.~M. Krauss, \rm Phys.~Rev.~Lett. \bf
     64\rm, 999 (1990).

\bibitem{haberkane} H.~E.~Haber and G.~L.~Kane, \rm
     Phys. Rep. \bf 117\rm, 75 (1985).

\bibitem{sneutrino} T. Falk, K. A. Olive, and M. Srednicki,
     Phys. Lett. {\bf B339}, 248 (1994).

\bibitem{ellishag} J.~Ellis, et al., \rm Nucl.~Phys. \bf
     B238\rm, 453 (1984); K. Griest, M. Kamionkowski and
     M. S. Turner, {\rm Phys.~Rev.} {\bf D41}, 3565 (1990). 

\bibitem{witten} M.~W.~Goodman and E.~Witten, Phys. Rev. \bf
     D31, \rm 3059 (1986); I. Wasserman, {\rm Phys.~Rev.} {\bf
     D33}, 2071 (1986); K.~Freese, J.~Frieman, and A.~Gould,
     Phys. Rev. \bf D37, \rm 3388 (1988); K.~Griest, \rm
     Phys. Rev. \bf D38\rm, 2357 (1988); FERMILAB-Pub-89/139-A
     (E); A.~Drukier, K.~Freese, and D.~Spergel, Phys. Rev. \bf
     D33, \rm 3495 (1986).

\bibitem{labdetectors} For a review of current detector
     development, see, J. Low Temp. Phys. \bf 93 \rm (1993). 

\bibitem{SOS} J.~Silk, K.~Olive, and M.~Srednicki,
     Phys.~Rev.~Lett. {\bf 55}, 257 (1985); K.~Freese, {\rm
     Phys. Lett.} {\bf B167}, 295 (1986); L.~M.~Krauss,
     K.~Freese, D.~N.~Spergel, and W. H. Press, {\rm
     Astrophys. J.} {\bf 299}, 1001 (1985); L. M.~Krauss,
     M.~Srednicki, and F.~Wilczek, {\rm Phys.~Rev.} {\bf D33},
     2079 (1986); T.~Gaisser, G.~Steigman, and
     S.~Tilav, {\rm Phys.~Rev.} {\bf D34}, 2206 (1986);
     M.~Kamionkowski, {\rm Phys.~Rev.} {\bf D44}, 3021 (1991);
     Ritz and D. Seckel, Nucl. Phys. {\bf B304}, 877 (1988);
     F. Halzen, M. Kamionkowski, and T. Stelzer, Phys. Rev.
     {\bf D45}, 4439 (1992).

\bibitem{imb} J.~M.~LoSecco et al.\ (IMB Collaboration), \rm Phys.~Lett.
	 \bf B188\rm, 388 (1987).

\bibitem{joakim} L. Bergstrom, J. Edsj\"o, and M. Kamionkowski, in
     preparation. 

\bibitem{wonyong} W. Lee, private communication.

\bibitem{cdms} P. D. Barnes et al., J. Low Temp. Phys. {\bf 93}, 79
     (1993); T. Shutt et al., Phys. Rev. Lett. {\bf 61}, 3425 (1992);
     ibid. 3531 (1992).

\bibitem{taorich} J. Rich and C. Tao, DAPNIA/SPP 95-01.

\bibitem{bernard} M. Kamionkowski, et al., Phys. Rev. Lett. {\bf
     74}, 5174 (1995).  

\bibitem{kanewells} G. L. Kane and J. D. Wells,
     Phys. Rev. Lett. {\bf 76}, 4458 (1996); S. Ambrosanio et
     al., Phys. Rev. Lett {\bf 76}, 3498 (1996).

\bibitem{CDF} S. Park, ``Search for New Phenomena at CDF,'' 10th
     Topical Workshop on Proton-Antiproton Collider Physics,
     edited by R. Raja and J. Yoh (AIP Press, 1996).

\bibitem{katie} K. Freese and M. Kamionkowski, hep-ph/9609370.

\bibitem{kim} K.~Griest, \rm Phys. Rev. \bf D38\rm, 2357 (1988);
     FERMILAB-Pub-89/139-A (E).

\bibitem{drees} M. Drees and M. M. Nojiri, Phys. Rev. {\bf
     D47}, 4226 (1993); ibid. {\bf 48}, 3483 (1993).

\bibitem{MAP} http://map.gsfc.nasa.gov. 

\bibitem{COBRAS} http://www.estec.esa.nl/spdwww/future/html/cosa.htm.

\bibitem{kamspergelsug} M.~Kamionkowski, D.~N.~Spergel, and N.~Sugiyama,
     {\rm Astrophys.~J.~Lett.} {\bf 426}, L57 (1994).

\bibitem{jkksone} G. Jungman, M. Kamionkowski, A. Kosowsky, and
     D. N. Spergel, Phys. Rev. Lett. {\bf 76}, 1007 (1996).

\bibitem{jkkstwo} G. Jungman, M. Kamionkowski, A. Kosowsky, and
     D. N. Spergel, Phys. Rev. {\bf D54}, 1332 (1996).

\bibitem{gds} S. Dodelson, E. Gates, and A. Stebbins,
     Astrophys. J. {\bf 467}, 10 (1996).

\bibitem{polarization} M. Kamionkowski, A. Kosowsky, and
     A. Stebbins, astro-ph/9609132; U. Seljak and
     M. Zaldarriaga, astro-ph/9609169


}
\end{thebibliography}
\end{document}